\documentclass[a4paper,11pt]{article}
\pdfoutput=1 
\usepackage{jheppub} 

\usepackage{tikz}

	\newcommand{\comments}[1]{}
	
	\newcommand\be{\begin{equation}}
		\newcommand\ee{\end{equation}}
	\newcommand\bea{\begin{align}}
		\newcommand\eea{\end{align}

}
	
	\def\beqa{\begin{eqnarray}}
\def\eeqa{\end{eqnarray}}

\title{
Celestial Amplitudes: Conformal Partial Waves and Soft Limits
}

\author{Dhritiman Nandan,$^1$}
\author{Anders Schreiber,$^{2}$}
\author{Anastasia Volovich,$^{2}$}
\author{Michael Zlotnikov$^3$}

\affiliation{$^1$ School of Physics \& Astronomy, 
University of Southampton,
Highfield, Southampton, SO17 1BJ,
United Kingdom.}

\affiliation{$^2$ Department of Physics,
  Brown University,
  Providence RI 02912.}

\affiliation{$^3$ Center for Theoretical Physics,
Department of Physics,
Colombia University,
New York, NY 10027.}

\emailAdd{d.nandan@soton.ac.uk}
\emailAdd{anders\_schreiber@brown.edu}
\emailAdd{anastasia\_volovich@brown.edu}
\emailAdd{mz2737@columbia.edu}

\abstract{\\
Massless scattering amplitudes in four-dimensional Minkowski spacetime can be Mellin-transformed to correlation functions on the celestial sphere at null infinity called celestial amplitudes.
We study various properties of massless four-point scalar and gluon celestial amplitudes 
such as conformal partial wave 
decomposition, crossing relations and optical theorem. As a byproduct, 
we derive the analog of the single and double soft limits for all gluon celestial amplitudes.
}

\begin{document} 

\maketitle

\flushbottom

\section{Introduction}

Pasterski, Shao, and Strominger (PSS) have
  proposed a map between $\mathcal{S}$-matrix elements in four-dimensional Minkowski spacetime 
and correlation functions in two-dimensional conformal field theory (CFT) living on the celestial sphere~\cite{Pasterski:2016qvg,Pasterski:2017ylz}.
Celestial CFT is interesting both for understanding the long elusive holographic description of flat spacetime 
\cite{flat}
as well as for exploring the mathematical structures of amplitudes. 
In recent years many remarkable properties of amplitudes have been uncovered via twistor space, momentum twistor space, 
scattering equations, etc.(see~\cite{Elvang:2013cua}~for review),
hence it is quite plausible that exploring properties of celestial amplitudes may also
lead to new insights.

A key idea behind the PSS proposal was to transform the plane wave basis to a manifestly conformally covariant basis
called the conformal primary wavefunction basis.
This basis was constructed explicitly by Pasterski and Shao \cite{Pasterski:2017kqt} for particles of various spins
in diverse dimensions. The celestial sphere is the null infinity of  four-dimensional Minkowski spacetime. The 
double cover of the
four-dimensional Lorentz group 
is identified with the $SL(2,\mathbb{C})$ conformal group of the celestial sphere.
Two-dimensional correlators on the celestial sphere will be referred to as celestial amplitudes from here on.

The celestial amplitudes of \emph{massless} particles are given by Mellin transforms
of the corresponding four-dimensional amplitudes
\begin{align}
	\label{map}
	\mathcal{\tilde A}_n(z_j, \bar{z}_j)= \int_0^\infty \prod_{l=1}^{n}  d\omega_l\,\omega_l^{\Delta_l-1} 
	\mathcal{A}_n(k_l) \, ,
\end{align}
where  $\Delta_l = 1 + i \lambda_l$, with $\lambda_l \in \mathbb{R}$ \cite{Pasterski:2017kqt}, are conformal dimensions taking values in the principal continuous series, in order to ensure the orthogonality
and completeness of the conformal primary wavefunction basis.  Further details are given below.

In the spirit of recent developments in understanding scattering amplitudes from the on-shell perspective by studying symmetries, analytic properties, and unitarity,
 many recent studies have delved into similar aspects of celestial amplitudes.
The structure of factorization of singularities of celestial amplitudes was investigated in \cite{Cardona:2017keg}; 
three- and four-point gluon amplitudes were computed in \cite{Pasterski:2017ylz} and arbitrary tree-level ones in \cite{Schreiber:2017jsr}.
Celestial four-point string amplitudes have been discussed in \cite{Stieberger:2018edy}. Unitarity via the manifestation of the optical theorem on celestial amplitudes has been observed recently \cite{Banerjee:2017jeg,Lam:2017ofc}
and the generators of Poincar\'e and conformal groups in the celestial representation were constructed in
\cite{Stieberger:2018onx}.

This paper is organized as follows.
In section~\ref{scal4massless4} we  compute massless scalar four-point celestial amplitudes and study its properties
such as conformal partial wave decomposition, crossing relations, and optical theorem.
In section~\ref{exampleGluon} we derive conformal partial wave decomposition for four-point gluon celestial amplitude, 
and in section~\ref{soft} single and double soft limits for all gluon celestial amplitudes. The conformal partial wave decomposition formalism is summarized in appendix~\ref{CPWappendix} and details about inner product integrals, required in the main text, are evaluated in appendix~\ref{InnPr}.
\smallskip

Note added: During this work, we became aware of related work by Pate, Raclariu, and Strominger \cite{Pate:2019} which has some overlap with section 4 of our paper.

\section{Scalar Four-Point Amplitude}
\label{scal4massless4}
\begin{figure}
\vspace{-0.9in}
\centering
\begin{tikzpicture}[line width=1.7 pt,baseline={([yshift=-3ex]current bounding box.center)},vertex/.style={anchor=base,
    circle,fill=black!25,minimum size=18pt,inner sep=2pt}]
\draw[style] (-1.05, 0) -- (0.55, 0);
\draw[dashed] (-2.1, 0.75) --(-1.05, 0)  ;
\draw[dashed]  (-2.1, -0.75)-- (-1.05, 0);
\draw[dashed] (0.55, 0) -- (1.6, 0.75);
\draw[dashed]  (0.55, 0)-- (1.6, -0.75);
\node[scale=1] at (0, 3.3) {$~$};
\node[scale=1] at (-0.2, 0.35) {$m$};
\node[scale=1] at (-2.5, 0.75) {$k_2$};
\node[scale=1] at (-2.5, -0.75) {$k_1$};
\node[scale=1] at (2.0, 0.75) {$k_3$};
\node[scale=1] at (2.0, -0.75) {$k_4$};
\draw[dashed] (3.4, 0.75) -- (4.75, 0.5);
\draw[dashed] (4.75, 0.5) -- (6.1, 0.75);
\draw[dashed] (3.4, -0.75) -- (4.75, -0.5);
\draw[dashed] (4.75, -0.5) -- (6.1, -0.75);
\draw[style] (4.75, -0.5) -- (4.75, 0.5);
\node[scale=1] at (3, 0.75) {$k_2$};
\node[scale=1] at (3, -0.75) {$k_1$};
\node[scale=1] at (6.5, 0.75) {$k_3$};
\node[scale=1] at (6.5, -0.75) {$k_4$};
\node[scale=1] at (5.15, 0) {$m$};
\draw[dashed] (8, 0.75) -- (9.35, 0.5);
\draw[dashed] (9.35, 0.5) -- (10.7, -0.75);
\draw[dashed] (8, -0.75) -- (9.35, -0.5);
\draw[dashed] (9.35, -0.5) -- (10.7, 0.75);
\draw[style] (9.35, -0.5) -- (9.35, 0.5);
\node[scale=1] at (7.6, 0.75) {$k_2$};
\node[scale=1] at (7.6, -0.75) {$k_1$};
\node[scale=1] at (11.1, 0.75) {$k_3$};
\node[scale=1] at (11.1, -0.75) {$k_4$};
\node[scale=1] at (8.95, 0) {$m$};
\end{tikzpicture}
\caption{Four-Point Exchange Diagrams} \label{fig:00M00}
\end{figure}
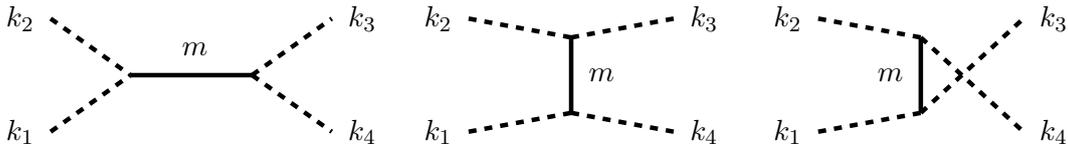

In this section we study a tree level four-point amplitude of massless  scalars 
mediated by exchange of 
 a massive scalar depicted on Figure \ref{fig:00M00}\footnote{The same amplitude in three dimensions was studied in \cite{Lam:2017ofc}.}. 

The corresponding celestial amplitude (\ref{map}) is
\begin{align}
\label{fourmassless}
\mathcal{\tilde A}_4(z_j, \bar{z}_j)= g^2\int_0^\infty \prod_{j=1}^4 d\omega_j\,\omega_j^{\Delta_j-1} 
\delta^{(4)} \left(\sum_{i=1}^4 k_i \right)\left(\frac{1}{(k_1{+} k_2)^2{+}m^2}{+}\frac{1}{(k_1{+} k_3)^2{+}m^2}{+}\frac{1}{(k_1{+} k_4)^2{+}m^2}\right) \, ,
\end{align}
where $z_j,\bar z_j$ are coordinates on the celestial sphere and $\omega_j$ are the energies. 
Defining $\epsilon_{j} =- 1$ (+1) for incoming (outgoing) particles, we can parameterize the momenta
$k^\mu_j$ as
\begin{align}
	k^\mu_j=\epsilon_j \omega_j \left(1+|z_j|^2,z_j+\bar z_j,i\bar z_j-iz_j,1-|z_j|^2\right) \, .
	\label{k-pss}
\end{align}

Under conformal transformations,  by construction \cite{Pasterski:2017kqt}, the four-point celestial amplitude behaves as a four-point CFT correlation function of operators with conformal weights
\begin{align}
(h_j ,  \bar{h}_j ) =\frac{1}{2}(\Delta_j+J_j , \Delta_j -J_j)  \, ,
\end{align} 
where $J_j$ are spins. We can split the four-point celestial amplitude into a conformally invariant function of only the cross-ratios $ \tilde A_4(z,\bar z) $ and a universal prefactor 
\begin{align}
	\label{basisExpansion}
	\mathcal{\tilde A}_{4}(z_j , \bar z_j) = \frac{\left(\frac{z_{24}}{z_{14}}\right)^{h_{12}}\left(\frac{z_{14}}{z_{13}}\right)^{h_{34}}}{z_{12}^{h_1+h_2}z_{34}^{h_3+h_4}}\frac{\left(\frac{\bar z_{24}}{\bar z_{14}}\right)^{\bar h_{12}}\left(\frac{\bar z_{14}}{\bar z_{13}}\right)^{\bar h_{34}}}{\bar z_{12}^{\bar h_1+\bar h_2}\bar z_{34}^{\bar h_3+\bar h_4}} \tilde A_4(z,\bar z) \, ,
\end{align}
where we define $h_{ij}=h_i-h_j$, $\bar h_{ij}=\bar h_i-\bar h_j$ and cross-ratios
\begin{align}
	\label{ratios}
	z=\frac{z_{12}z_{34}}{z_{13}z_{24}} \, , ~~\bar{z}=\frac{\bar{z}_{12}\bar{z}_{34}}{\bar{z}_{13}\bar{z}_{24}}~~~\text{with}~~~z_{ij}=z_i-z_j \, ,
	~~\bar{z}_{ij}=\bar{z}_i-\bar{z}_j \, .
\end{align}

Let's fix the external points in \eqref{fourmassless} as $z_1=0, z_2=z, z_3=1, z_4=1/\tau$ with $\tau\to 0$, and compute
\begin{align}
\label{prescription}
\tilde A_{4}(z) \equiv|z|^{\Delta_1+\Delta_2} \lim_{\tau\to 0} \tau ^{-2 \Delta _4}\mathcal{\tilde A}_{4}(0,z,1,1/\tau) \, .
\end{align}

We will consider the case where particles 1 and 2 are incoming while 3 and 4 are outgoing, so $\epsilon_{1}=\epsilon_{2}=-\epsilon_{3}=-\epsilon_{4}=-1$ and denote it as
 $12 \leftrightarrow 34$. 
 The $s$-channel diagram on figure \ref{fig:00M00} is
\begin{align}
\label{schannel}
\tilde A_{4,s}^{12 \leftrightarrow 34} (z) \sim g^2 |z|^{\Delta_1+\Delta_2} \lim_{\tau\to 0} \tau ^{-2 \Delta _4}\int_0^\infty \prod_{i=1}^4 d\omega_i\,\omega_i^{\Delta_i-1} 
\delta^{(4)}\left(\sum_{j=1}^4 k_j \right)\frac{1}{m^2-4   \omega_1\omega_2 |z|^2} \, .
\end{align}
The momentum conservation delta functions can be rewritten as:
\begin{align}
\label{threedeltas}
\delta^{(4)}\left(\sum_{j=1}^4 k_j \right)=\frac{4\tau^2}{ \omega_1}\delta(i\bar z-iz)\prod_{i=2}^4\delta(\omega_i-\omega_i^*) \, ,
\end{align}
where  
\begin{align}
\label{Udef}
\omega_2^*=\frac{\omega_1}{ z-1}~~,~~\omega_3^*=\frac{z \omega_1}{z-1}~~,~~\omega_4^*=z \omega_1 \tau^2 \, .
\end{align}
The delta function only has solutions when all the $\omega_i^*$ are positive, so $z>1$. 

Then  \eqref{schannel} reduces to a single integral
\begin{align}  \notag
\tilde A_{4,s}^{12 \leftrightarrow 34} (z) &\sim  g^2 \delta(i\bar z-iz) z^{\Delta_1+\Delta_2}  \lim_{\tau\to 0} \tau ^{2-2 \Delta _4}\int_0^\infty d\omega_1 \omega_1^{\Delta_1-2}
\,\prod_{i=2}^4 (\omega^*_i)^{\Delta_i-1}\frac{1}{m^2-   \frac{4 z^2}{z-1}\omega_1^2}\\
&= \frac{g^2\left(i m/2\right)^{2 \alpha-2}}{\sin({\pi} \alpha)}  \delta(i\bar z-iz)\,z^2\,(z-1)^{h_{12}-h_{34}} \, . \label{scalarCase}
\end{align}
Adding the $s$-, $t$-, and $u$-channel contributions, 
we obtain our final result
\begin{align}
\label{chanSscal}
\tilde A_{4}^{12 \leftrightarrow 34} (z)\sim g^2
\frac{ ({m/2})^{2 \alpha-2} }{\sin(\pi \alpha)}  
\delta(i\bar z-iz)\,z^2\,(z-1)^{h_{12} - h_{34} } 
\left( e^{\pi i\alpha}+
\left(\frac{z}{z-1}\right)^\alpha+z^\alpha\right) \, ,
\end{align}
where 
\begin{align}
\alpha=\sum_{i=1}^4 h_i -2 \, .
\end{align}

Let us discuss some properties of this expression.

\smallskip

$\bullet$ First, it is straightforward to verify that the Poincar\'e generators on the celestial sphere, 
constructed in \cite{Stieberger:2018onx},
\begin{align} 
\begin{split}
L_{1,i} &= (1- z_i^2) \partial_{z_i} - 2 z_i h_i \, , ~~ \\
\bar L_{1,i} &= (1- \bar z_i^2) \partial_{\bar z_i} - 2 \bar z_i \bar h_i  \, , ~~ \\
P_{0,i} &= (1+ |z_i|^2) e^{(\partial_{h_i} + \partial_{\bar{h}_i} )/2} \, ,  \\
P_{2,i} &= - i (z_i -  \bar{z}_i) e^{(\partial_{h_i} + \partial_{\bar{h}_i} )/2} \, , 
\end{split} 
\begin{split}
L_{2, i} &= (1+z_i^2) \partial_{z_i} + 2 z_i h_i  \, , ~~~~ L_{3, i}  = 2 (z_i \partial_{z_i} + h_i) \, , \\
\bar L_{2, i} &= (1+ \bar z_i^2) \partial_{\bar z_i} + 2 \bar z_i \bar h_i \, , ~~~~ \bar L_{3, i}  = 2 (\bar z_i \partial_{\bar z_i} + \bar h_i) \, , \\
P_{1,i} &= (z_i + \bar{z}_i) e^{(\partial_{h_i} + \partial_{\bar{h}_i } )/2} \, , \\
P_{3,i} &= (1- |z_i|^2) e^{(\partial_{h_i} + \partial_{\bar{h}_i} )/2}
\end{split}
\end{align}
annihilate the celestial amplitude on the support of the delta function $\delta(i\bar z-i z)$.

\smallskip

$\bullet$ Second, we can show that $\tilde A_{4}$ satisfies the crossing relations
\begin{align}
\label{cross1}
\tilde A_{4}^{13 \leftrightarrow 24} (1-z)=\left(\frac{1-z}{z}\right)^{2 (h_2 + h_3) }\tilde A_{4}^{13 \leftrightarrow 24} (z)\,,~~~~~~0<z<1\,,
\end{align}
as well as
\begin{align}
\label{cross2}
\begin{split}
\tilde A_{4}^{13 \leftrightarrow 24} (z)&= z^{2 (h_1 + h_4 )} \tilde A_{4}^{12 \leftrightarrow 34} (1/z)\\
&=(1-z)^{2(h_{12}-h_{34})} \tilde A_{4}^{14 \leftrightarrow 23} \left(\frac{z}{z-1}\right)
\end{split}\,,~~~~~~0<z<1\,.
\end{align}
The relations \eqref{cross1} and \eqref{cross2} generalize similar relations in \cite{Lam:2017ofc}.

\smallskip

$\bullet$ Third,  the conformal partial wave decomposition
of $s$-channel celestial amplitude \eqref{scalarCase}\footnote{The other two channels can be obtained in similar manner.} is computed in the appendix~\ref{CPWappendix} and takes the following form
\begin{align}
\label{psiImfssum}
 \tilde A_{4,s}^{12 \leftrightarrow 34}(z)&\sim \frac{g^2\left(im/2\right)^{2\alpha-2}}{2\sin(\pi \alpha)}
\int_{{\cal{C}}} \frac{d\Delta}{4\pi^2}\,\frac{\Gamma \left(1{-}\frac{\Delta }{2}{-}h_{12}\right) \Gamma \left(\frac{\Delta
   }{2}{-}h_{12}\right) \Gamma \left(1{-}\frac{\Delta }{2}{-}h_{34}\right) \Gamma \left(\frac{\Delta
   }{2}{-}h_{34}\right)}{\Gamma (1-\Delta ) \Gamma (\Delta-1 )} {\Psi^{\Delta}_{h_i}(z,\bar z)} \, ,
\end{align}
where ${\Psi^{\Delta}_{h_i}(z,\bar z)}$ is given in (\ref{psibasis}) restricted to the internal scalar case with $J=0$ and the contour ${\cal{C}}$ runs from $1 - i \infty$ to $1 + i \infty$.

The gamma functions in (\ref{psiImfssum}) unambiguously specify all pole sequences in conformal dimensions. 
Closing the contour to the right or left of the complex axis in $\Delta$, we find simple poles  at 
$\Delta$ and their shadows at  $\tilde{\Delta}$ given by
\begin{align}
\frac{ \Delta }{2}=1-h_{12}+n~~~,~~~\frac{\Delta }{2}=1-h_{34}+n ~~~ , ~~~\frac{\tilde \Delta}{2}=h_{12}-n~~~,~~~\frac{\tilde \Delta}{2}=h_{34}-n  \, ,
\end{align}
with $n=0,1,2,3,...\,$. 

$\bullet$ Finally, let's explicitly check the celestial optical theorem derived by
Shao and Lam in \cite{Lam:2017ofc}, which
 relates the imaginary part of the four-point celestial amplitude to
the product of two three-point celestial amplitudes with the appropriate integration measure.
Taking imaginary part of  \eqref{psiImfssum} we obtain 
\begin{align}
\text{Im}\left[ \tilde A_{4,s}^{12 \leftrightarrow 34} (z)\right]\sim\int_{\mathcal{C}} d \Delta \,\mu( \Delta )C( h_1, h_2; \Delta)C( h_3, h_4; 2 - \Delta){\boldsymbol{\Psi}^{\Delta}_{h_i} (z,\bar z)} \, ,
\end{align}
up to some overall constants independent of $h_i$. Here $C(h_i,h_j;\Delta)$ is the coefficient
of the three-point function
given by \cite{Lam:2017ofc}
\begin{align}
C(  h_i,  h_j; \Delta) = g \frac{(m^2)^{ h_i + h_j -2 }}{4^{h_i +  h_j} } \frac{\Gamma \left(  h_{ij} + \frac{\Delta}{2}\right) \Gamma \left( \frac{\Delta}{2} - h_{ij}  \right)}{\Gamma(\Delta)} \, ,
\end{align}
 $\mu(\Delta)$ is the integration measure
\begin{align}
\mu (\Delta) = \frac{\Gamma(\Delta) \Gamma(2 - \Delta)}{4 \pi^3 \Gamma(\Delta-1) \Gamma(1 - \Delta)} \, ,
\end{align}
and $\boldsymbol{\Psi}^{\Delta}_{h_i} (z,\bar z)$ is 
\begin{align}
\boldsymbol{\Psi}^{\Delta}_{h_i} (z,\bar z)\equiv\frac{\Gamma \left(1-\frac{\Delta }{2}-h_{12}\right) \Gamma \left(\frac{\Delta
   }{2}-h_{34}\right)}{\Gamma \left(\frac{\Delta }{2}+h_{12}\right) \Gamma \left(1-\frac{\Delta
   }{2}+h_{34}\right)}\Psi^{\Delta}_{h_i}(z,\bar z) \, .
\end{align}

\section{Gluon Four-Point Amplitude}
\label{exampleGluon}
In this section we study the massless four-point gluon celestial amplitude, which has 
 been computed in  \cite{Pasterski:2017ylz}, and is given by  
 \begin{align}
\label{fglu}
\tilde A_{--++}^{12 \leftrightarrow 34}(z)\sim \delta(i\bar z-iz)|z|^{3}|1-z|^{h_{12}-h_{34}-1} \, , ~~~ z> 1 \, ,
\end{align}
where the conformal ratios $z,\bar z$ are defined in (\ref{ratios}). 

Evaluating the integral in appendix \ref{InnPr}, we find the conformal partial wave expansion is given by the following simple result\footnote{When considering $J<0$, take $h\leftrightarrow \bar h$ in the expansion coefficient.}
\begin{align}
\tilde A_{--++}^{12 \leftrightarrow 34}(z) \sim 
i {\sum_{J=0}^{\infty}}'\int_{\cal{C}} \frac{d\Delta }{4\pi^2}\, \Psi^{h,\bar h}_{h_i,\bar h_i}\frac{\pi \, (1-2h)(2\bar h-1)}{(h_{34}{-}h_{12})\sin(\pi (h_{12}{-}h_{34}) )} \left(\frac{\Gamma(h{-}h_{12})\Gamma(1{-}h_{34}{-}\bar h)}{\Gamma(h{+}h_{12})\Gamma(1{+}h_{34}{-}\bar h)}{+}(h_{12}\leftrightarrow h_{34})\right) \, ,
\end{align}
where $\sum'$ means that the $J = 0$ term contributes with weight $1/2$. 

There is no truncation of the spins $J$ in this case, so primary operators of all integer spins contribute to the OPE expansion of the external gluon operators, in contrast with the previously considered scalar case.

Poles $\Delta$ and shadow poles $\tilde{\Delta}$ are located at
\begin{align}
 \frac{\Delta -J}{2}=1-h_{12}+n~~~,~~~ \frac{ \Delta -J}{2}=1-h_{34}+n ~~~, ~~~ \frac{\tilde \Delta+J}{2}=h_{12}-n~~,~~\frac{\tilde \Delta+J}{2}=h_{34}-n \, ,
\end{align}
with $n=0,1,2,3,...\,$. These poles are  integer spaced as expected.

\section{Soft limits}
\label{soft}

\noindent\textit{Single soft limits} 
\par
In this section we study the analog of soft limits for celestial amplitudes. The universal soft behavior of color-ordered gluon scattering amplitudes, corresponding to $\omega_k \rightarrow 0$, is well-known \cite{Weinberg:1965nx} and takes the form
\begin{align} \label{amplitudesoft}
\begin{split}
\lim_{\omega_k \rightarrow 0} \mathcal{A}_n^{\ell_k = +1} &= \frac{\langle k -1 \, k+1 \rangle}{\langle k-1 \, k \rangle \langle k \, k+1 \rangle} \mathcal{A}_{n-1} \,  , \\
\lim_{\omega_k \rightarrow 0} \mathcal{A}_n^{\ell_k = -1} &= \frac{[ k -1 \, k+1 ]}{[ k-1 \, k ][ k \, k+1 ]} \mathcal{A}_{n-1} \, , 
\end{split}
\end{align}
where $\ell_k$ is the helicity of particle $k$.

The spinor-helicity variables are related to the celestial sphere variables via \cite{Pasterski:2017ylz}
\begin{align}
\label{spihel}
[ij] =  2\sqrt{\omega_i \omega_j} \bar z_{ij} \,  ,~~~~~\langle ij\rangle = -2 \epsilon_i \epsilon_j\sqrt{\omega_i \omega_j } z_{ij} \, .
\end{align}
Conformal primary wavefunctions become soft (pure gauge) when $\Delta_k \rightarrow 1$ (or $\lambda_k \rightarrow 0$) \cite{Pasterski:2017kqt, Donnay:2018neh}. In this limit we can utilize the delta function representation\footnote{See \url{http://mathworld.wolfram.com/DeltaFunction.html}}
\begin{align} \label{deltarep}
\delta (x) = \frac{1}{2} \lim_{\lambda \rightarrow 0}  \, i \lambda  \, |x|^{i \lambda-1} \, ,
\end{align}
such that \eqref{map} becomes
\begin{align} \label{mapsoft}
\lim_{\lambda_k \rightarrow 0} \mathcal{\tilde A}_n (z_j, \bar z_j) =
\frac{1}{i \lambda_k}  \prod_{j=1, j \ne k}^n  \int_0^\infty d\omega_j \, \omega_j^{i\lambda_j } \, \int_0^\infty d\omega_k \, 2 \, \delta(\omega_k) \, \omega_k  \, \mathcal{A}_n (\omega_j; z_j,\bar z_j) \, .
\end{align}
We see that the $\lambda_k \rightarrow 0$ limit localizes the integral at $\omega_k = 0$ and we obtain
\begin{align}\label{celestialcorrsoft1}
\lim_{\lambda_k \rightarrow 0} \mathcal{\tilde A}_n^{J_k = +1} &= \frac{1}{i \lambda_k} \frac{z_{k-1 \, k+1}}{z_{k-1 \, k} z_{k \, k+1} }  \mathcal{\tilde A}_{n-1}  \, , \\
\lim_{\lambda_k \rightarrow 0} \mathcal{\tilde A}_n^{J_k = -1} &= \frac{1}{i \lambda_k} \frac{\bar z_{k-1 \, k+1}}{\bar z_{k-1 \, k} \bar z_{k \, k+1} }  \mathcal{\tilde A}_{n-1} \, . \label{celestialcorrsoft2}
\end{align}
These relations were checked for four- and six-point MHV celestial amplitudes in \cite{Fan:2019emx}.

\vspace{11pt}
\noindent\textit{Double soft limits} 
\par
For consecutive soft limits, one can apply \eqref{celestialcorrsoft1} or \eqref{celestialcorrsoft2} multiple times and the consecutive soft factors are simply products of single soft factors.

For simultaneous double soft limits, energies of particles are simultaneously scaled by $\delta$, so $\omega_k \rightarrow \delta \omega_k$ and $\omega_l \rightarrow \delta \omega_l$ with $\delta \rightarrow 0$, which for example yields \cite{Volovich:2015yoa, Klose:2015xoa}
\begin{align} \label{doublesoftamp}
\lim_{\delta \rightarrow 0} \mathcal{A} _n (\delta \omega_1 , \delta \omega_2, \omega_j; z_k, \bar z_k) = \frac{1}{\langle n | 1 +2 | 3] } \left( \frac{[13]^3 \langle n 3 \rangle}{[12 ] [23] s_{123} } + \frac{\langle n2 \rangle^3 [n 3] }{\langle n 1 \rangle \langle 1 2 \rangle s_{n12} } \right) \mathcal{A} _{n-2} ( \omega_j; z_j, \bar z_j)  \, , 
\end{align}
for $\ell_1 = +1$, $\ell_2 = -1$, $j = 3, \ldots, n$, and $k = 1, \ldots, n$. Here $s_{ijl} = (k_i + k_j + k_l)^2$. More generally we will write
\begin{align} \label{doublesoftampgen}
\lim_{\delta \rightarrow 0} \mathcal{A} _n (\delta \omega_k , \delta \omega_l, \omega_j; z_i, \bar z_i) =  \text{DS} (k^{\ell_k} , l^{\ell_l}) \mathcal{A} _{n-2} ( \omega_j; z_j, \bar z_j)  \, ,
\end{align}
where $\text{DS} (k^{\ell_k} , l^{\ell_l})$ is the simultaneous double soft factor.

For celestial amplitudes, the analog of the simultaneous double soft limit is to take two $\lambda$'s, scale them by $\epsilon$, $\lambda_k \rightarrow \epsilon\lambda_k$ and $\lambda_l \rightarrow \epsilon \lambda_l$, and take the $\epsilon\rightarrow 0$ limit. To implement this practically in \eqref{map}, we change variables for the associated $\omega$'s
\begin{align} \label{doublesoftcoords}
\omega_k = r \cos(\theta), ~~ \omega_l = r \sin (\theta), ~~ 0\leq r < \infty, ~~ 0 \leq \theta \leq \frac{\pi}{2} \, .
\end{align}
The mapping \eqref{map} becomes
\begin{align} \label{simdoublesoft2}
\begin{split}
\mathcal{\tilde A}_{n} (z_j, \bar z_j)
&=  \prod_{j=1, j \ne k , l}^n  \int_0^\infty d\omega_j \, \omega_j^{i\lambda_j } \, \int_0^\infty dr   \, \int_0^{\pi/2}  d\theta \, r^{( i \lambda_k + i \lambda_l) \epsilon  -1} \\
&\times (\cos(\theta))^{i\lambda_k \epsilon} (\sin (\theta))^{i\lambda_l \epsilon} r^2 \mathcal{A}_{n} (\omega_j; z_j,\bar z_j)\, .
\end{split}
\end{align}
We can use \eqref{deltarep} to obtain a delta function in $r$, which enforces the simultaneous double soft limit for the scattering amplitude as in \eqref{doublesoftamp}. The result is
\begin{align}\label{corrsimdoubsoft}
\lim_{\epsilon \rightarrow 0}  \mathcal{\tilde A}_n  (\lambda_k \epsilon, \lambda_l \epsilon) = \tilde{\text{DS}} (k^{J_k}, l^{J_l})  \mathcal{\tilde A}_{n-2} \, ,
\end{align}
where $ \tilde{\text{DS}} (k^{J_k}, l^{J_l}) $ is the simultaneous double soft factor on the celestial sphere,
\begin{align} \label{celesdoublesoftfac}
 \tilde{\text{DS}} (k^{J_k}, l^{J_l})  =  \frac{1}{ (i \lambda_k + i \lambda_l) \epsilon}   \, \left[  2 \int_0^{\pi/2}  d\theta \,   (\cos(\theta))^{i\lambda_k \epsilon} (\sin (\theta))^{i\lambda_l \epsilon} \left[  r^2 \text{DS} (k^{\ell_k}, l^{\ell_l}) \right]_{r = 0}  \right]_{\epsilon \rightarrow 0} \, .
\end{align}
As an example, consider the simultaneous double soft factor in \eqref{doublesoftamp}. We can use \eqref{spihel} to translate it into celestial sphere coordinates and plug into \eqref{celesdoublesoftfac} to obtain
\begin{align}
 \tilde{\text{DS}} (1^{+1}, 2^{-1})  \sim \frac{1}{ 2  (i \lambda_1 + i \lambda_2)  \epsilon^2 } \frac{1}{z_{n1} \bar{z}_{23}} \left(    \frac{1}{i\lambda_1}  \frac{ \bar{z}_{n3} z_{2n}   }{z_{12} \bar{z}_{2n}} + \frac{1}{i \lambda_2} \frac{ z_{3n}  \bar{z}_{31} }{\bar{z}_{12} z_{31}}  \right) \, .
\end{align}

It is straightforward to generalize \eqref{corrsimdoubsoft} to $m$ particles taken simultaneously soft, by introducing $m$-dimensional spherical coordinates as in \eqref{doublesoftcoords} and scale $m$ $\lambda$'s by $\epsilon$.

\acknowledgments
We are grateful to L. Donnay, J. Drummond, D. Kapec, D. Lowe, S. Pasterski, M. Pate, A.-M. Raclariu, S.-H. Shao, M. Spradlin, A. Strominger, and C. Tan for useful conversations and C. Wen for collaboration on the
earlier stages of this project. DN is supported by ERC grant 648630 IQFT. This work is supported in part by the US Department of Energy under contract DE-SC0010010 Task A (AS, AV); and by Simons Investigator Award \#376208 (AS, AV). MZ is supported by the US Department of Energy under contract DE-SC0011941.

\appendix

\newpage

\section{Conformal Partial Wave Decomposition}
\label{CPWappendix}

In the CFT four-point function defined as \eqref{basisExpansion}, we can expand the conformally invariant part $\tilde A_4 (z,\bar z)$ on the basis of conformal partial waves $\Psi^{h,\bar h}_{h_i,\bar h_i}(z,\bar z)$. 
As can be shown along the lines of \cite{Caron-Huot:2017vep, Murugan:2017eto, Simmons-Duffin:2017nub}, the expansion takes the following form
\begin{align}
\label{decomposition}
\tilde A_4 (z,\bar z)= i {\sum_{J=0}^{\infty}}' \int_{\mathcal{C}} d \Delta \, \Psi^{h,\bar h}_{h_i,\bar h_i}(z,\bar z)\frac{(1-2h)(2\bar h-1)}{(2\pi)^2}\langle  \tilde A_4 (z,\bar z),\Psi^{h,\bar h}_{h_i,\bar h_i}(z,\bar z) \rangle \, ,
\end{align}
where  $h-\bar{h}=J$, $h+\bar h=\Delta=1+i \lambda$. The contour ${\cal{C}}$ runs from $1 - i \infty$ to $1 + i \infty$.
The integration and summation is over all dimensions and spins of exchanged primary operators in the theory.
${\sum}'$ means that the $J=0$ summand contributes with a weight of $1/2$.
The inner product is defined by
\begin{align}
\label{innerpr}
\langle G(z,\bar z),F(z,\bar z) \rangle \equiv\int \frac{dz d\bar z}{(z\bar z)^2}G(z,\bar z)\overline {F(z,\bar z)} \, .
\end{align}

 The  conformal partial waves $\Psi^{h,\bar h}_{h_i,\bar h_i}(z,\bar z)$ have been computed in \cite{Dolan:2003hv, Dolan:2011dv, Osborn:2012vt} and are given by 
\begin{align}
\label{psibasis}
\Psi^{h,\bar h}_{h_i,\bar h_i}(z,\bar z)=&c'_1F_{+}(z,\bar z)+c'_2F_{-}(z,\bar z) \, ,
\end{align}
with
\begin{align}
F_{+}(z,\bar z)=&\frac{1}{z^{h_{34}}\bar{z}^{\bar{h}_{34}}}  \, _2F_1\left({{1-h+h_{34},h+h_{34}}\atop{1+h_{12}+h_{34}}};\frac{1}{z}\right) \,
   _2F_1\left({{1-\bar{h}+\bar{h}_{34},\bar{h}+\bar{h}_{34}}\atop{1+\bar{h}_{12}+\bar{h}_{34}}};\frac{1}{\bar{z}}\right) \, ,\\
F_{-}(z,\bar z)=&z^{h_{12}} \bar{z}^{\bar{h}_{12}} \, _2F_1\left({{1-h-h_{12},h-h_{12}}\atop{1-h_{12}-h_{34}}};\frac{1}{z}\right) \,
   _2F_1\left({{1-\bar{h}-\bar{h}_{12},\bar{h}-\bar{h}_{12}}\atop{1-\bar{h}_{12}-\bar{h}_{34}}};\frac{1}{\bar{z}}\right) \, ,\notag\\
c'_1=&(-1)^{h-\bar{h}+h_{12}-\bar{h}_{12}}\frac{\Gamma \left(-h_{12}-h_{34}\right)}{\Gamma \left(1+\bar{h}_{12}+\bar{h}_{34}\right)}\frac{ \Gamma \left(1-h+h_{12}\right)  \Gamma \left(h+h_{34}\right) \Gamma
   \left(\bar{h}+\bar{h}_{12}\right) \Gamma \left(1-\bar{h}+\bar{h}_{34}\right)}{\Gamma \left(1-h-h_{12}\right) \Gamma \left(h-h_{34}\right) \Gamma
   \left(\bar{h}-\bar{h}_{12}\right) \Gamma \left(1-\bar{h}-\bar{h}_{34}\right) } \, ,\notag\\
c'_2=&(-1)^{h-\bar{h}+h_{34}-\bar{h}_{34}}\frac{ \Gamma \left(h_{12}+h_{34}\right)}{\Gamma \left(1-\bar{h}_{12}-\bar{h}_{34}\right)} \, .\notag
\end{align}
Here we made use of hypergeometric identities discussed in \cite{Dolan:2011dv} to rewrite the result in a form which is suited for the region $z, \bar{z} >1$.

Conformal partial waves are orthogonal with respect to the
inner product (\ref{innerpr})
\begin{align}
\langle \Psi^{h,\bar h}_{h_i,\bar h_i}(z,\bar z),\Psi^{h',\bar h'}_{h_i,\bar h_i}(z,\bar z) \rangle =\frac{(2\pi)^2}{(1-2h)(2\bar h-1)}\delta_{J,J'}\delta(\lambda-\lambda') \, .
\end{align}

The basis functions (\ref{psibasis}) span a complete basis for bosonic fields on each of the ranges
\begin{align}
(J\in\mathbb{Z}~,~\lambda\in\mathbb{R}^+~|~J\in\mathbb{Z}^+~,~\lambda\in\mathbb{R}~|~J\in\mathbb{Z}~,~\lambda\in\mathbb{R}^-~|~J\in\mathbb{Z}^-~,~\lambda\in\mathbb{R}) \, .
\end{align}

We can perform the $\Delta$ integration in (\ref{decomposition}) by collecting residues of poles located at "physical" exchanged operator dimension values. One can use e.g. the integral representation of the conformal partial wave (\ref{psibasis}) (given by eq. (7) in \cite{Osborn:2012vt}) to make sure that the half-circle integration at infinity vanishes.

\section{Inner Product Integral }
\label{InnPr}

In this appendix we evaluate the inner product
\begin{align} \label{innerProdInt}
\langle \tilde A_4(z, \bar{z}), \Psi^{h,\bar h}_{h_i,\bar h_i}(z,\bar z) \rangle \equiv\int \frac{dz d\bar z}{(z\bar z)^2}\delta(i\bar z-iz)\,|z|^{2+\sigma}\,|z-1|^{h_{12}-h_{34}-\sigma}\,\overline { \Psi^{h,\bar h}_{h_i,\bar h_i}(z,\bar z)} \, ,
\end{align}
for $\sigma=0$ and $\sigma=1$, where $\Psi^{h,\bar h}_{h_i,\bar h_i}(z,\bar z)$ is given by \eqref{psibasis}\footnote{Note that in both of our examples we have $\bar h_{ij}=h_{ij}$, and the complex conjugation prescription $h\to 1-\bar h$, $\bar h\to 1-h$, $h_{ij}\to - h_{ij}$ and $z\leftrightarrow \bar z$}.

First we change integration variables to $z=x+i y $, $\bar z=x-i y$ and localize the delta function on $y=0$. Subsequently, we write the hypergeometric functions from (\ref{psibasis}) in the following Mellin-Barnes representation:
\begin{align}
\label{MellBarn}
{}_2F_1(a,b;c;z)=\frac{\Gamma (c)}{\Gamma (a) \Gamma (b) \Gamma (c-a) \Gamma (c-b)}\int_C\frac{ds}{2\pi i}(1-z)^s \Gamma (-s)\Gamma (c-a-b-s) \Gamma (a+s) \Gamma (b+s)  \, ,
\end{align}
where $(1-z)\in\mathbb{C}\backslash\mathbb{R_-}$ and the contour $C$ goes from minus to plus complex infinity while separating pole sequences in $ \Gamma (-s)\Gamma (c-a-b-s) $ from pole sequences in $\Gamma (a+s) \Gamma (b+s)$. 

The $x>1$ integral then gives a beta function which we express in terms of gamma functions. At this point, similarly to section 3.4 in \cite{Hogervorst:2017sfd}, the gamma function arguments in the integrand arrange themselves exactly such that one of the Mellin-Barnes integrals (\ref{MellBarn}) can be evaluated by second Barnes lemma.\footnote{We assume the integrals to be regulated appropriately such that these formal manipulations hold.} The final inverse Mellin transform integral is then done by closing the integration contour to the left, or to the right of the complex axis. Performing the sum over all residues of poles wrapped by the contour in this process we obtain
\begin{align}
\label{innerPrRes}
&\langle  \tilde A_4 (z,\bar{z}) , \Psi^{h,\bar h}_{h_i,\bar h_i}(z,\bar z) \rangle ~=~  \pi ^2 (-1)^{h-\bar{h}} \csc \left(\pi  \left(h_{12}-h_{34}\right)\right) \csc \left(\pi 
   \left(h_{12}+h_{34}\right)\right) \Gamma (1-\sigma )\\
	&\left[\left( \frac{\Gamma \left(1-\sigma +h_{12}-h_{34}\right) \, _4F_3\left({{1-\sigma
   ,1-\bar{h}+h_{12},\bar{h}+h_{12},1-\sigma +h_{12}-h_{34}}\atop{2-h-\sigma +h_{12},h-\sigma
   +h_{12}+1,h_{12}-h_{34}+1}};1\right)}{\Gamma \left(h_{12}-h_{34}+1\right) \Gamma
   \left(1-\bar{h}+h_{34}\right) \Gamma \left(\bar{h}+h_{34}\right) \Gamma \left(2-h-\sigma
   +h_{12}\right) \Gamma \left(h-\sigma +h_{12}+1\right)}-(h_{12}\leftrightarrow h_{34})\right) \right.\notag\\
	&+\left.\frac{\left(\frac{\Gamma \left(1-\bar{h}-h_{12}\right) \Gamma \left(\bar{h}-h_{12}\right) \Gamma \left(1-\sigma
   -h_{12}+h_{34}\right) }{\Gamma
   \left(1-h_{12}+h_{34}\right) \Gamma \left(2-h-\sigma -h_{12}\right) \Gamma \left(h-\sigma
   -h_{12}+1\right)}\, _4F_3\left({{1-\sigma ,1-\bar{h}-h_{12},\bar{h}-h_{12},1-\sigma
   -h_{12}+h_{34}}\atop{2-h-\sigma -h_{12},h-\sigma -h_{12}+1,1-h_{12}+h_{34}}};1\right)-(h_{12}\leftrightarrow h_{34})\right)}{\Gamma \left(1-h+h_{12}\right) \Gamma \left(h+h_{12}\right) \Gamma \left(1-h+h_{34}\right)
   \Gamma \left(h+h_{34}\right)}\right] \, , \notag
\end{align}
where we used identities such as $\sin(x+\pi \bar h)\sin(y+\pi \bar h)=\sin(x+\pi  h)\sin(y+\pi  h)$ for integer $J$, and $\sin(\pi x)=\pi/(\Gamma(x)\Gamma(1-x))$ to write \eqref{innerPrRes} in a shorter form.

\vspace{11pt}
\noindent\textit{Evaluation for $\sigma = 0$} 
\par

When $\sigma=0$, one upper and one lower parameter in the $\,_4F_3$ hypergeometric functions become equal and cancel, so that the functions reduce to $\,_3F_2$. Interestingly, an even greater simplification occurs as
\begin{align}
\, _3F_2\left({{1,a-c+1,a+c}\atop{a-b+2,a+b+1}};1\right)=\frac{\frac{\Gamma (a-b+2) \Gamma (a+b+1)}{\Gamma (a-c+1) \Gamma (a+c)}-(a-b+1) (a+b)}{(b-c) (b+c-1)} \, .
\end{align}
Then, making use of various sine- and gamma function identities as mentioned above, it turns out that the result is proportional to
\begin{align}
\frac{\sin(2\pi J)}{2\pi J}=\left\{{{1~;J=0}\atop{0~;J\neq 0}}\right. \, .
\end{align}
Therefore, the only non-vanishing inner product in this case comes from the scalar conformal partial wave $\Psi^{\Delta}_{h_i}\equiv \left. \Psi^{h,\bar h}_{h_i,\bar h_i} \right|_{J=0}$, which simplifies to
\begin{align}
\label{scalInnPr}
&\langle  \tilde A_4 (z, \bar{z}) , \Psi^{\Delta}_{h_i}(z,\bar z) \rangle = \frac{\Gamma \left(1-\frac{\Delta }{2}-h_{12}\right) \Gamma \left(\frac{\Delta
   }{2}-h_{12}\right) \Gamma \left(1-\frac{\Delta }{2}-h_{34}\right) \Gamma \left(\frac{\Delta
   }{2}-h_{34}\right)}{  \Gamma (2-\Delta ) \Gamma (\Delta )} \, .
\end{align}

\vspace{11pt}
\noindent\textit{Evaluation for $\sigma =1$} 
\par

As we take $\sigma\to 1$ the overall factor $\Gamma (1-\sigma )$ diverges. However, the rest of the terms conspire to cancel this pole, so that the limit $\sigma\to 1$ is finite. The simplification of the result in all generality is quite tedious, here we instead discuss a less rigorous but quick way to arrive at the end result.

The cases for the first few values of $J=0,1,...$ can be simplified directly e.g. in Mathematica. We recognize that the result is always proportional to $\csc(\pi (h_{12}-h_{34}) )/(h_{12}-h_{34})$. To quickly arrive at the full result, start with (\ref{innerPrRes}) and divide out the overall factor $\csc(\pi (h_{12}-h_{34})) / (h_{12} - h_{34})$. By the previous observation we see that the rest is finite in $h_{12}-h_{34}\to 0$. Sending $h_{34} \to h_{12}$ under a small $1-\sigma$ deformation, the hypergeometric functions become equal to $1$ for $\sigma\to 1$ and the remaining terms simplify. To recover the full $h_{12},h_{34}$ dependence it then suffices to match these terms e.g. to the specific example in the case $J=1$, which then for all $J\geq 0$ leads to

\begin{align}
\label{gluInnPr}
\langle  \tilde A_4 (z, \bar{z}) , \Psi^{h,\bar h}_{h_i,\bar h_i}(z,\bar z) \rangle=\frac{\pi\csc(\pi (h_{12}-h_{34}))}{(h_{34}-h_{12})} \left(\frac{\Gamma(h-h_{12})\Gamma(1-h_{34}-\bar h)}{\Gamma(h+h_{12})\Gamma(1+h_{34}-\bar h)}+(h_{12}\leftrightarrow h_{34})\right) \, .
\end{align}
To obtain the result for $J<0$ substitute $h\leftrightarrow \bar h$.


\begin{thebibliography}{99}


\bibitem{Pasterski:2016qvg} 
  S.~Pasterski, S.~H.~Shao and A.~Strominger,
  ``Flat Space Amplitudes and Conformal Symmetry of the Celestial Sphere,''
  arXiv:1701.00049 [hep-th].
  
  
\bibitem{Pasterski:2017ylz} 
  S.~Pasterski, S.~H.~Shao and A.~Strominger,
  ``Gluon Amplitudes as 2d Conformal Correlators,''
  Phys.\ Rev.\ D {\bf 96}, no. 8, 085006 (2017)
  doi:10.1103/PhysRevD.96.085006
  [arXiv:1706.03917 [hep-th]].
  
 \bibitem{flat} 
  J.~de Boer and S.~N.~Solodukhin,
  ``A Holographic reduction of Minkowski space-time,''
  Nucl.\ Phys.\ B {\bf 665}, 545 (2003)
  doi:10.1016/S0550-3213(03)00494-2
  [hep-th/0303006].
	 G.~Barnich and C.~Troessaert,
  ``Symmetries of asymptotically flat 4 dimensional spacetimes at null infinity revisited,''
  Phys.\ Rev.\ Lett.\  {\bf 105}, 111103 (2010)
  doi:10.1103/PhysRevLett.105.111103
  [arXiv:0909.2617 [gr-qc]].
   T.~Banks,
  ``The Super BMS Algebra, Scattering and Holography,''
  arXiv:1403.3420 [hep-th].
   A. ~Ashtekar, ``Asymptotic Quantization: Based On 1984 Naples
Lectures,`` Naples, Italy: Bibliopolis,(1987).
    A.~Strominger,
  ``Lectures on the Infrared Structure of Gravity and Gauge Theory,''
  arXiv:1703.05448 [hep-th].
	C.~Cheung, A.~de la Fuente and R.~Sundrum,
  ``4D scattering amplitudes and asymptotic symmetries from 2D CFT,''
  JHEP {\bf 1701}, 112 (2017)
  doi:10.1007/JHEP01(2017)112
  [arXiv:1609.00732 [hep-th]].
  
  
   
  
  \bibitem{Elvang:2013cua} 
  H.~Elvang and Y.~t.~Huang,
  ``Scattering Amplitudes,''
  arXiv:1308.1697 [hep-th].
  
\bibitem{Pasterski:2017kqt} 
  S.~Pasterski and S.~H.~Shao,
  ``Conformal basis for flat space amplitudes,''
  Phys.\ Rev.\ D {\bf 96}, no. 6, 065022 (2017)
  doi:10.1103/PhysRevD.96.065022
  [arXiv:1705.01027 [hep-th]].


  


\bibitem{Cardona:2017keg} 
  C.~Cardona and Y.~t.~Huang,
  ``S-matrix singularities and CFT correlation functions,''
  JHEP {\bf 1708}, 133 (2017)
  doi:10.1007/JHEP08(2017)133
  [arXiv:1702.03283 [hep-th]].

	

\bibitem{Schreiber:2017jsr} 
  A.~Schreiber, A.~Volovich and M.~Zlotnikov,
  ``Tree-level gluon amplitudes on the celestial sphere,''
  Phys.\ Lett.\ B {\bf 781}, 349 (2018)
  doi:10.1016/j.physletb.2018.04.010
  [arXiv:1711.08435 [hep-th]].



		
\bibitem{Stieberger:2018edy} 
  S.~Stieberger and T.~R.~Taylor,
  ``Strings on Celestial Sphere,''
  arXiv:1806.05688 [hep-th].
 

\bibitem{Banerjee:2017jeg} 
  N.~Banerjee, S.~Banerjee, S.~Atul Bhatkar and S.~Jain,
  ``Conformal Structure of Massless Scalar Amplitudes Beyond Tree level,''
  JHEP {\bf 1804}, 039 (2018)
  doi:10.1007/JHEP04(2018)039
  [arXiv:1711.06690 [hep-th]].
	


\bibitem{Lam:2017ofc} 
  H.~T.~Lam and S.~H.~Shao,
  ``Conformal Basis, Optical Theorem, and the Bulk Point Singularity,''
  arXiv:1711.06138 [hep-th].

 
  \bibitem{Stieberger:2018onx} 
  S.~Stieberger and T.~R.~Taylor,
  ``Symmetries of Celestial Amplitudes,''
  arXiv:1812.01080 [hep-th].
	

\bibitem{Pate:2019}
  M.~Pate, A.~M.~Raclariu, and A.~Strominger,
  ``Conformally Soft Theorem in Gauge Theory,''
  to appear. 

\bibitem{Weinberg:1965nx} 
  S.~Weinberg,
  ``Infrared photons and gravitons,''
  Phys.\ Rev.\  {\bf 140}, B516 (1965).
  doi:10.1103/PhysRev.140.B516

\bibitem{Donnay:2018neh} 
  L.~Donnay, A.~Puhm and A.~Strominger,
  ``Conformally Soft Photons and Gravitons,''
  JHEP {\bf 1901}, 184 (2019)
  doi:10.1007/JHEP01(2019)184
  [arXiv:1810.05219 [hep-th]].

\bibitem{Fan:2019emx} 
  W.~Fan, A.~Fotopoulos and T.~R.~Taylor,
  ``Soft Limits of Yang-Mills Amplitudes and Conformal Correlators,''
  arXiv:1903.01676 [hep-th].
	


\bibitem{Volovich:2015yoa} 
  A.~Volovich, C.~Wen and M.~Zlotnikov,
  JHEP {\bf 1507}, 095 (2015)
  doi:10.1007/JHEP07(2015)095
  [arXiv:1504.05559 [hep-th]].



\bibitem{Klose:2015xoa} 
  T.~Klose, T.~McLoughlin, D.~Nandan, J.~Plefka and G.~Travaglini,
  ``Double-Soft Limits of Gluons and Gravitons,''
  JHEP {\bf 1507}, 135 (2015)
  doi:10.1007/JHEP07(2015)135
  [arXiv:1504.05558 [hep-th]].



\bibitem{Caron-Huot:2017vep} 
  S.~Caron-Huot,
  ``Analyticity in Spin in Conformal Theories,''
  JHEP {\bf 1709}, 078 (2017)
  doi:10.1007/JHEP09(2017)078
  [arXiv:1703.00278 [hep-th]].

\bibitem{Simmons-Duffin:2017nub} 
  D.~Simmons-Duffin, D.~Stanford and E.~Witten,
  ``A spacetime derivation of the Lorentzian OPE inversion formula,''
  arXiv:1711.03816 [hep-th].
	
\bibitem{Murugan:2017eto} 
  J.~Murugan, D.~Stanford and E.~Witten,
  ``More on Supersymmetric and 2d Analogs of the SYK Model,''
  JHEP {\bf 1708}, 146 (2017)
  doi:10.1007/JHEP08(2017)146
  [arXiv:1706.05362 [hep-th]].
	


\bibitem{Dolan:2003hv} 
  F.~A.~Dolan and H.~Osborn,
  ``Conformal partial waves and the operator product expansion,''
  Nucl.\ Phys.\ B {\bf 678}, 491 (2004)
  doi:10.1016/j.nuclphysb.2003.11.016
  [hep-th/0309180].

\bibitem{Dolan:2011dv} 
  F.~A.~Dolan and H.~Osborn,
  ``Conformal Partial Waves: Further Mathematical Results,''
  arXiv:1108.6194 [hep-th].

\bibitem{Osborn:2012vt} 
  H.~Osborn,
  ``Conformal Blocks for Arbitrary Spins in Two Dimensions,''
  Phys.\ Lett.\ B {\bf 718}, 169 (2012)
  doi:10.1016/j.physletb.2012.09.045
  [arXiv:1205.1941 [hep-th]].

 
  	
\bibitem{Hogervorst:2017sfd} 
  M.~Hogervorst and B.~C.~van Rees,
  ``Crossing symmetry in alpha space,''
  JHEP {\bf 1711}, 193 (2017)
  doi:10.1007/JHEP11(2017)193
  [arXiv:1702.08471 [hep-th]].




 
  \end{thebibliography}
\end{document}